# Superconductivity in three-layer $Na_{0.3}CoO_2 \cdot 1.3H_2O$

**M.L. Foo[1], T. Klimczuk[1,3], Lu Li[2], N.P. Ong[2], and R.J. Cava[1]**

[1] Department of Chemistry, Princeton University, Princeton NJ 08544,

[2] Department of Physics, Princeton University, Princeton NJ 08544,

[3] Faculty of Applied Physics and Mathematics, Gdansk University of Technology,

Narutowicza 11/12, 80-952 Gdansk, Poland,

The observation of superconductivity at 4.3 K in a new crystalline form of $Na_{0.3}CoO_2 \cdot 1.3H_2O$ is reported. The new superconductor has three layers of $CoO_6$ octahedra per crystallographic unit cell, in contrast to the previously reported two-layer superconductor. The three-layer cell occurs because the relative orientations of neighboring $CoO_2$ layers are distinctly different from what is seen in the two-layer superconducting phase. This type of structural difference in materials that are otherwise chemically and structurally identical is not possible to attain on the layered copper oxide superconductors. The synthesis and stability of the new phase are described.

Since the discovery of superconductivity near 4 K in $Na_{0.35}CoO_2 \cdot 1.35H_2O$[1], the physics of electronic and magnetic systems based on layered triangular lattices, where the geometry frustrates long-range magnetic ordering at low temperatures, has been of great interest. The chemistry and structure of $Na_{0.35}CoO_2 \cdot 1.35H_2O$, which consists of two triangular cobalt oxide sheets per unit cell separated by spacer layers of water molecules and sodium ions, as well as its apparent balance between magnetism and superconductivity, are similar to the cuprate high Tc superconductors. Here we report the discovery of superconductivity in three-layer $Na_{0.30}CoO_2 \cdot 1.3H_2O$, at a temperature of 4.3 K. Although essentially identical in chemical composition to the original two-layer superconductor, this superconductor is structurally distinct due to differences in the stacking of the $CoO_2$ sheets in the third dimension. The consequences of this kind of structural difference have not been pursued in the cuprate superconductors. The observation of superconductivity at the same critical temperature in both two-layer and three-layer sodium cobalt oxyhydrates suggests that the essential physics of these superconductors will be well described by theoretical models that consider the magnetic and electronic system to be essentially two-dimensional in nature.

The parent phase of the original $Na_{0.35}CoO_2 \cdot 1.35H_2O$ superconductor is $Na_{0.7}CoO_2$, which is obtained by heating sodium and cobalt oxides at 800 degrees in oxygen. The unit cell is built up by two (2) sheets of edge-shared $CoO_6$ octahedra, which are rotated by 60 degrees with respect to each other. The sodium ions, on two crystallographically distinct sites, are trigonal prismatic (P) in coordination to the oxygen atoms, and hence the nomenclature P2 (prismatic, two layer) is used to describe the structure of $Na_{0.7}CoO_2$. A distinctly different thermodynamic form of sodium cobalt oxide, $NaCoO_2$, is obtained by heating sodium and cobalt oxides at 500 degrees in oxygen[2]. $NaCoO_2$, which is isostructural with $\alpha$-$NaFeO_2$, has three (3) sheets of edge shared $CoO_6$ octahedra per unit cell, displaced laterally from each other. The sodium ions are only on one crystallographic site, in octahedral (O) coordination to the oxygen atoms, and hence the nomenclature O3 (octahedra, 3 layer) is used to describe the structure of $NaCoO_2$. It is this three-layer octahedrally coordinated Na phase that is the host compound for the new superconductor described here. The P2 and O3 structures of $Na_xCoO_2$ are compared in figure 1.

O3 $NaCoO_2$ was synthesized by an adaptation of the procedure used by Wiley and Cushing[3]. A 15% excess of NaOH pellets (EM Science, 97% minimum purity) and stiochiometric amounts of Co metal sponge (Johnson Matthey, Grade I) were mixed in a dense alumina crucible, followed by heating in $O_2$ at 500 degrees for 12 h. The fused mixture was ground and annealed at 800 degrees for 16 h in $N_2$ to yield a gray crystalline powder. The powder was then finely ground and treated with $Br_2$ concentration of 40x (1 x is the theoretical amount of $Br_2$ needed to remove all the Na in $NaCoO_2$) in $CH_3CN$ for 5 days. The product was washed in dry $CH_3CN$ and dried in an atmosphere of flowing argon. The Na content of the product was verified by neutron powder diffraction structural analysis and Inductive Coupled Plasma (ICP) analysis to give the formula $Na_{0.30}CoO_2$. The structural analysis, to be reported elsewhere[4], showed that the O3 structure type was fully maintained on deintercalation of the sodium. The bromine concentration employed to get the correct Na content to yield superconductivity seems to be more critical in the case of three-layer $Na_xCoO_2$ than in two-layer $Na_xCoO_2$.

Figure 1 shows powder X-ray Diffraction patterns (Cu K$\alpha$ radiation) of the three-layer compounds O3 $NaCoO_2$, $Na_{0.30}CoO_2$, and the "intermediate hydrate" of three-layer O3 $Na_{0.30}CoO_2$. The overall chemical behavior of the three-layer system on deintercalation of Na and intercalation of water is very similar to that observed in the two-layer sodium cobaltate. The $c$ lattice parameter of O3 $NaCoO_2$ expands on deintercalation of Na to $Na_{0.3}CoO_2$, as is also seen on deintercalation of two-layer P2 $Na_{0.7}CoO_2$. Similar to the case for P2 $Na_{0.7}CoO_2$, an intermediate hydrate is obtained in the O3 system by drying a water washed sample of $Na_{0.30}CoO_2$ at room temperature. In the intermediate hydrates (also called "monolayer hydrates"), the water molecules are in the same layer as the sodium ions, increasing the $c$-axis of $Na_{0.30}CoO_2$ due to their larger size. The new superconducting superhydrate (also called "bilayer hydrate", two layers of water molecules are found between the $CoO_2$ sheets) in the three-layer cobaltate was synthesized by exposing anhydrous O3 $Na_{0.30}CoO_2$ to water vapor in a humidity chamber with 100% relative humidity at room temperature for 1 to 6 days. It may also be synthesized from $Na_{0.3}CoO_2$ washed in water and subsequently exposed to water vapor in a 100% relative humidity chamber at room temperature. We have found the former method to be the more satisfactory of the two.

The hydration chemistry of the three-layer superconductor was studied by Thermogravimetric Analysis (TGA). The inset to figure 1 shows the change in weight for anhydrous three-layer $Na_{0.3}CoO_2$ on exposure to a stream of wet oxygen at approximately 40% relative humidity. The gain in weight represents the uptake of water. X-ray diffraction analysis of the product showed it to be the intermediate hydrate phase, yielding from the measured weight gain a formula for the three-layer intermediate hydrate of $Na_{0.3}CoO_2 \cdot 0.60H_2O$, in agreement with the formula proposed for the intermediate hydrate for the two-layer system[5]. Figure 2 shows the change in weight for the three-layer superconducting $Na_{0.3}CoO_2 \cdot xH_2O$ superhydrate on heating very slowly in flowing $O_2$. The behavior on heating is generally quite similar to that of the two-layer superconductor, but with significant differences. Water is lost in a series of steps on increasing temperature. The lowest temperature weight losses represent the evaporation of intergrain (surface) water, and the decomposition of the superhydrate to the intermediate hydrate. The plateau in weight from 35 to approximately $50^oC$ represents the stability region of the intermediate hydrate $Na_{0.3}CoO_2 \cdot 0.60H_2O$. This compound loses water in steps until decomposition to $Na_{0.5}CoO_2$ and $Co_3O_4$ at $300^oC$. The inset to figure 2 compares the details of the weight loss characteristics of the two-layer and three-layer superconducting superhydrates at low temperatures. The lowest temperature weight losses are very similar, representing the evaporation of surface water. The hump at intermediate temperatures represents the water loss on decomposition of the superhydrate to the intermediate hydrate. From the weight loss observed, and the known formula of the intermediate hydrate, the formula for the superconducting superhydrate can be determined to be $Na_{0.30}CoO_2 \cdot 1.3H_2O$ (in both cases). Looking carefully at the inset to figure 2 it is can be seen that that the three-layer superconducting superhydrate decomposes to the intermediate hydrate at 35 degrees, five degrees *lower* than the decomposition temperature of the two-layer superconductor, indicating that it is even less thermally stable than that phase, essentially barely stable at ambient temperature.

Extreme care must be taken in handling three-layer $Na_{0.30}CoO_2 \cdot 1.3H_2O$, as it decomposes fully to the intermediate hydrate after about 6 minutes exposure to 40% relative humidity air at ambient temperature. X-ray diffraction (XRD) characterization of three-layer $Na_{0.30}CoO_2 \cdot 1.3H_2O$ was therefore carried out in a sample holder that maintained the relative humidity around the sample at 100%. Figure 3 shows the XRD pattern of the three-layer superconductor compared to that of the two-layer superconductor. As the distances between the $CoO_2$ sheets of the two superconductors are almost equal, defining the crystallographic *c* axis, the first three (00l) peaks are found at the same diffracted angles for both phases. However, on examining the higher angle peaks (inset), clear differences are seen in the patterns, allowing unambiguous distinction between the two phases and the purity of the synthesized three-layer phase. The X-ray diffraction pattern can be indexed by the same R-3m space group as the parent O3 $NaCoO_2$ phase. Clearly, the room temperature chemistry performed on the three-layer O3 $NaCoO_2$ host sodium cobalt oxide to prepare the superhydrate superconductor does not provide sufficient energy to rotate the $CoO_6$ octahedra from their orientations in O3 $NaCoO_2$, allowing the synthesis of a superconducting compound with a distinctly different crystal structure from that of the two-layer form.

The superconductivity of the three-layer cobalt oxyhydrate was characterized by magnetization measurements (Quantum Design Squid Magnetometer), summarized in figure 4 for one of our preparations. The data show that the material is a bulk superconductor, with superconducting characteristics similar to that of the double-layer cobaltate. In the three-layer system, however, we reproducibly observed a dependence of the superconducting properties on length of hydration time. Figure 4 shows characteristic data illustrating this effect. In this set of experiments, anhydrous sodium cobaltate of nominal composition $Na_{0.3}CoO_2$, obtained in a 40X reaction with bromine, was placed in a hydration chamber where the hydration process is accomplished at ambient temperature on exposure to an atmosphere of 100% relative humidity. A small portion of the sample was removed in daily intervals for testing. The figure shows that the initial $T_c$ is in the range of 3K. The $T_c$ then increases to the optimal value, where it is stable for several days. Tc then degrades on further storage in the hydration atmosphere. Powder X-ray diffraction characterization of the material indicated that the three-layer $Na_{0.30}CoO_2 \cdot 1.3H_2O$ superconducting phase is present and single-phase to our experimental sensitivity for the duration of these experiments. Therefore the chemical changes that are taking place to yield the observations in figure 4 do not involve decomposition of the superconducting phase but rather subtle structural or chemical changes. One possibility is that some Na may be leached out of the superconductor into the surrounding intergrain water over a period of several days[6,7] an effect that is accelerated for hydration by immersion in large volumes of liquid water. The observed change in Tc may therefore represent the leaching of a slightly Na rich composition through the optimal composition to a Na poor composition over a period of a week, tracing out the narrow peak in Tc vs. Na content[8]. The lower chemical stability of the three-layer superhydrate at ambient conditions may be what allows this effect to be observed on a short laboratory time scale, but the general similarity of the chemistry of the phases suggests that this is may occur to a smaller degree in the two-layer superconductor as well.

We have demonstrated the existence of superconductivity in three-layer cobalt oxyhydrate $Na_{0.30}CoO_2 \cdot 1.3H_2O$, which has a distinctly different crystal structure from the two-layer sodium cobalt oxyhydrate. The difference in alignment of $CoO_2$ layers and oxygen coordination of the Na ions make the chemistry of the three-layer phase somewhat different from the two-layer phase, especially with respect to its stability in laboratory air at ambient temperature. This is likely due to differences in the $H_2O$ molecule arrangement necessary for bonding the two $H_2O$ layers to the oxygen in neighboring $CoO_2$ planes: neutron diffraction will be needed to shed light on the coordination of the water molecules and the Na ions in the three-layer superconductor. The identical superconducting critical temperatures observed in the two-layer and three-layer superconductors demonstrate that they are essentially electronically equivalent, and that electronic states that extend from one layer to the next do not appear to critically influence the superconductivity. The copper oxide superconductors do not include two compounds that differ only in the orientations or positions of neighboring layers while being identical in all other aspects of the chemistry and structure, such as formula and interlayer spacing, and so the effect of varying only the interlayer coupling can not be tested in the same manner as it can in the case of comparison of the properties of the two-layer and three-layer superconducting cobalt oxyhydrates. It will be of interest to compare the superconducting characteristics of these two phases in further detail.

## Acknowledgements

This research was supported by the National Science Foundation grants DMR 0244254 and DMR 0213706 and the Department of Energy, Basic Energy Sciences, grant DE FG02 98 ER45706. We gratefully acknowledge the contribution of Tao He at the Dupont Corporate Central Research and Development Laboratory.

**Figure Captions**

Figure 1: X-ray diffraction patterns (Cu Kα radiation) of a) O3 $NaCoO_2$, a=2.889(1) Å c=15.61(4) Å, b) $Na_{0.30}CoO_2$, a=2.8123(4) Å c=16.734(3) Å, c) and the monolayer hydrate $Na_{0.30}CoO_2 \cdot 0.6H_2O$, a=2.822(2) Å c= 20.81(1) Å. Lattice parameters of $Na_{0.30}CoO_2$ were obtained by neutron diffraction. Inset: The crystal structure of O3 $NaCoO_2$ and P2 $Na_{0.7}CoO_2$ with emphasis on the $CoO_6$ octahedra, gray spheres representing Na atoms. In O3 $NaCoO_2$, the Na atoms fully occupy one type of octahedrally coordinated site with oxygen. In the P2 structure of $Na_{0.7}CoO_2$, the Na ions are distributed over two types of triangular prismatic sites with partial occupancies.

Figure 2: Thermogravimetric analysis of the three-layer superconductor (3L SC) on heating in dry $O_2$ at $0.25^oC$/min showing the different hydrates obtained and decomposition products at temperatures up to 300 degrees. Inset: Detail of the low temperature region showing the different rate of weight loss for inter-grain and crystal water. Data from the two layer superconductor (2L SC) is included for comparison.

Figure 3: X-ray diffraction patterns of the two-layer (2L) and three-layer (3L) superhydrate superconductors. Inset: Enlargement of high angle region emphasizing the difference in crystal structures. Peaks are indexed using a $P6_3$/mmc space group for the 2L superconductor and an R-3m space group for the 3L superconductor.

Figure 4: Zero field cooled dc magnetization data, measured in a field of 5 Oe, for the three-layer sodium cobalt superhydrate superconductor showing the dependence of $T_c$ on the time of hydration. After day 4 the sample was stored in a refrigerator.

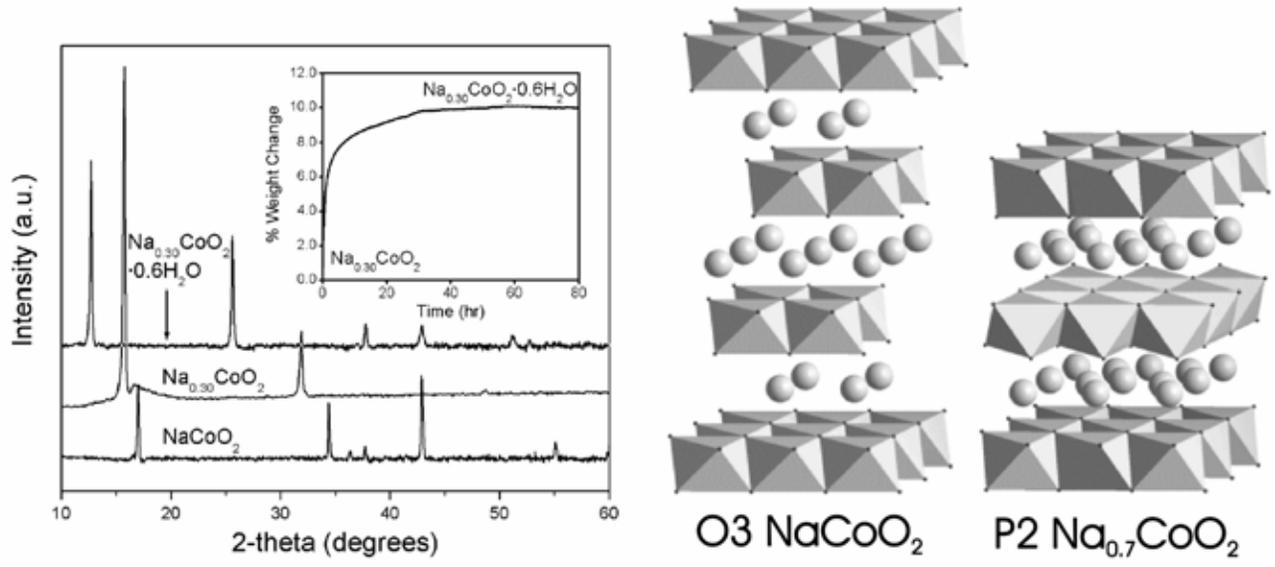

**Fig. 1**

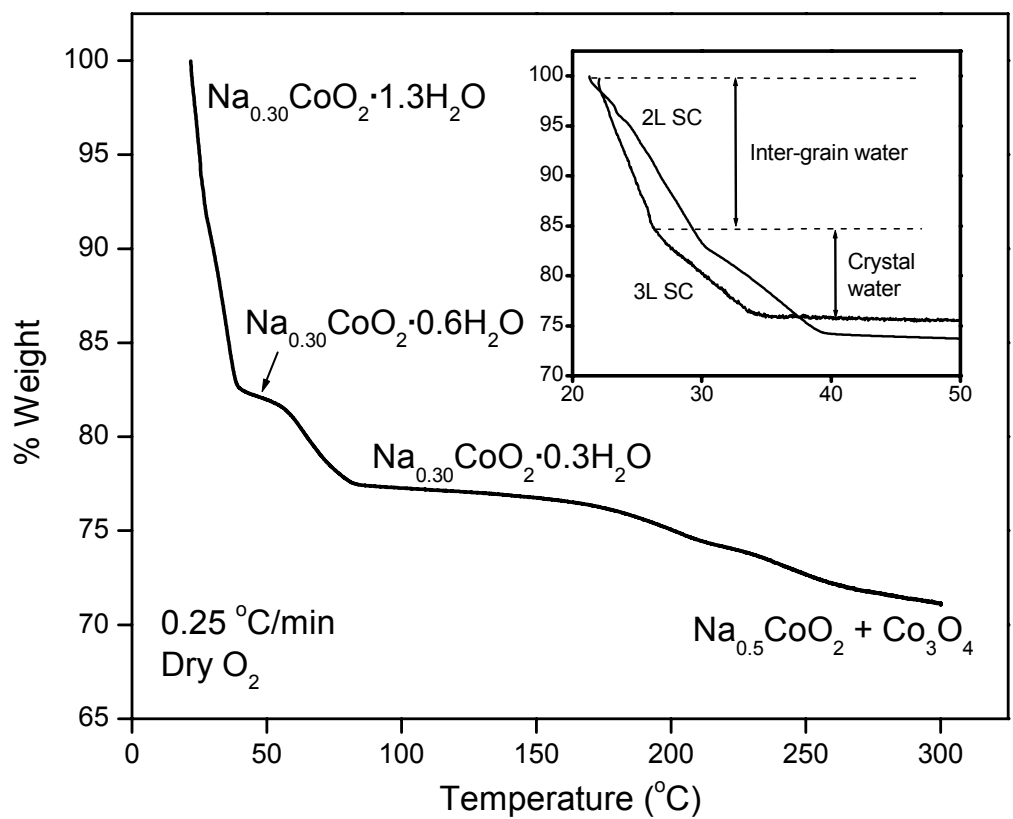

**Fig. 2**

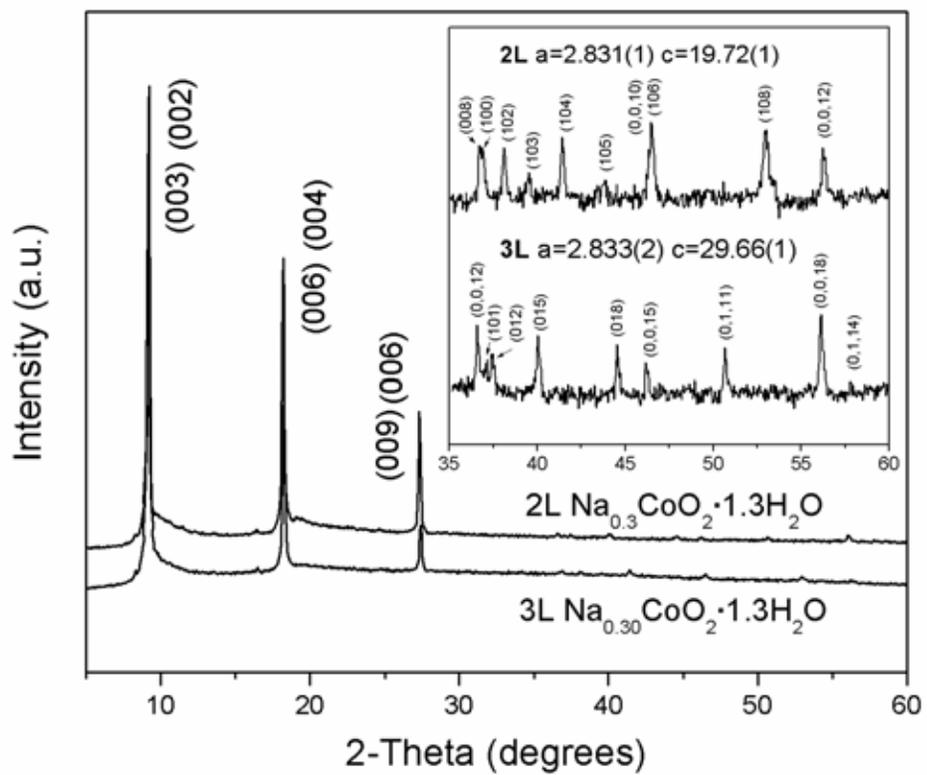

**Fig. 3**

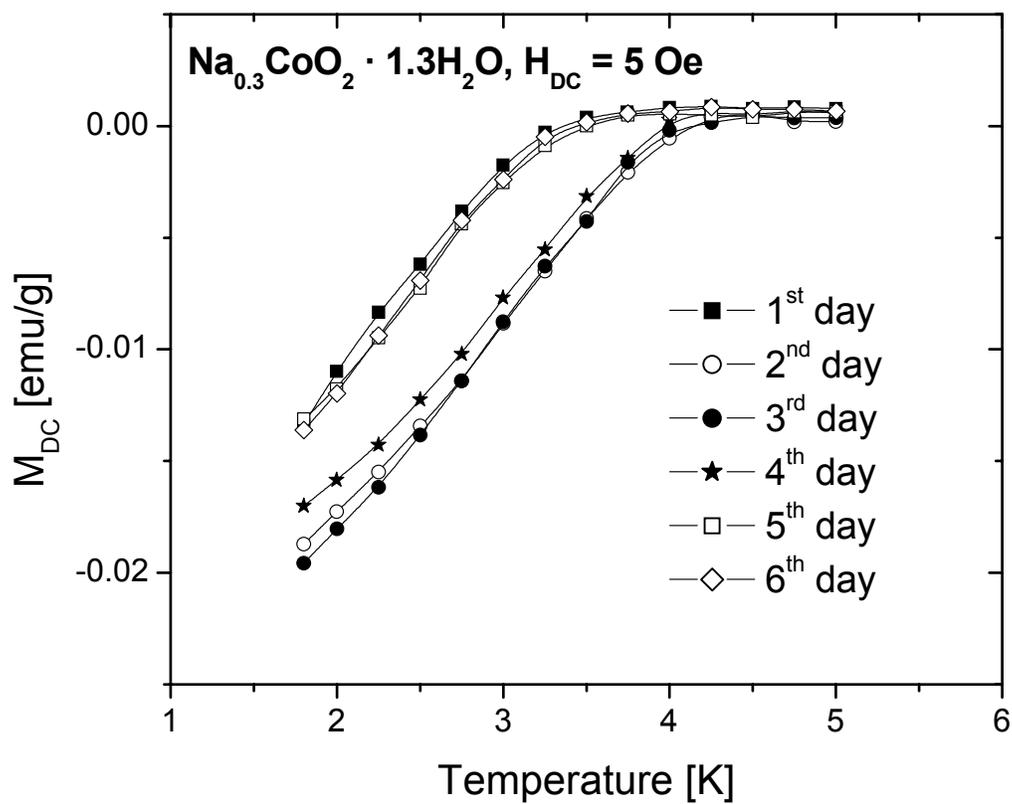

Fig. 4